\documentclass[twocolumn,showpacs,amsmath,amssymb,preprintnumbers,prb]{revtex4}

\usepackage{graphicx}
\usepackage{dcolumn}

\begin{document}


\title{Nanoengineering of a Negative-Index Binary-Staircase Lens for
the Optics Regime}
\author{B. D. F. Casse$^1$, R. K. Banyal$^1$, W. T. Lu$^{1,a)}$\footnote[16]{$^{a)}$ w.lu@neu.edu}}
\author{Y. J. Huang$^1$, S. Selvarasah$^2$, M. Dokmeci$^2$}
\author{S. Sridhar$^{1,b)}$\footnote[16]{$^{b)}$ s.sridhar@neu.edu}}
\affiliation{$^1$Department of Physics and
Electronic Materials Research Institute, Northeastern University,
Boston, Massachusetts 02115, USA} \affiliation{$^2$Department of
Electrical and Computer Engineering, Northeastern University,
Boston, Massachusetts 02115, USA}

\date{\today}

\begin{abstract}
We show that a binary-staircase optical element can be engineered to
exhibit an effective negative index of refraction, thereby expanding
the range of optical properties theoretically available for future
optoelectronic devices. The mechanism for achieving a negative-index
lens is based on exploiting the periodicity of the surface
corrugation. By designing and nanofabricating a plano-concave
binary-staircase lens in the InP/InGaAsP platform, we have
experimentally demonstrated at 1.55 $\mu$m that such
negative-index concave lenses can focus plane waves. The beam
propagation in the lens was studied experimentally and was in
excellent agreement with the three-dimensional finite-difference
time-domain numerical simulations.
\end{abstract}

\maketitle
%
The field of metaphotonics
\cite{shalaev_np07,veselago06,pendry96,pendry99,PhysRevB.62.10696,Gralak:00,luoprb02,casseapl07,gmachl07}, or the merging of
metamaterials \cite{veselago68} and photonics \cite{Joannopoulos95,yablonovitch87,sajeevjohn87}, has opened doors to a plethora
of unusual electromagnetic properties, such as negative refraction \cite{RAShelby04062001,parimi03}, cloaking
\cite{D.Schurig11102006} and optical data storage \cite{kosmas_nat_07}, that cannot be obtained with naturally occurring
materials. The holy grail of manufacturing these artificial photonic metamaterials structures, is to manipulate light at the
nanoscale level for optical information processing and high-resolution imaging. In this paper we demonstrate how a
binary-staircase optical element can be tailor-made to have an effective negative refractive index, and thus bringing an
alternative approach to negative-index optical elements.

Here we consider a binary-staircase type of lens \cite{alda05}, which consists of a
sequence of zones configured as flat parallel steps each having an
annular shape. The binary-staircase lens is a plano-concave lens.
Focusing by plano-concave lenses \cite{enoch03,parazzoli04} were
realized in 2D and 1D photonic crystals
\cite{vodo:201108,vodo:084104}. Proof-of-concept experiments which
demonstrate negative refraction in a plano-concave grating lens have
been realized earlier by our group in the microwave regime
\cite{luopt07}. However the plano-concave lens used in the microwave
range consisted of an assembly of commercial alumina bars, placed in
a parallel-plate waveguide, which are not suitable for integration
in optoelectronic circuits.

Geometrical parameters of the binary-staircase lens were determined
by considering the transverse size of the lens, the focal length,
wavelength of the incoming radiation, the index of the material used
to fabricate the lens itself and mainly the surface periodicity. The
actual lens has been nanofabricated by a combination of electron
beam lithography and reactive ion etching in an InP/InGaAsP
heterostructure. Subsequently, the focusing properties of the device
were experimentally verified using a scanning probe optical
technique. Three-dimensional (3D) finite-difference time-domain
(FDTD) simulations have been used to further study the beam
propagation in the lens. The FDTD simulations were in excellent
agreement with the experimental results.

\begin{figure}[htbp]
\centering{\includegraphics[width=5cm]{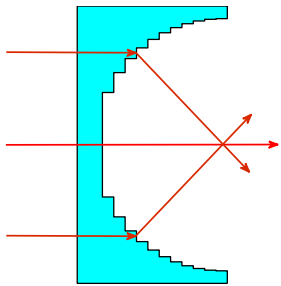}}
\caption{(Color online) Sketch of a plano-concave grating lens.
The lens is made of a medium with $n_{\mathrm{med}}$.
The horizontal step size $a$ is smaller than the free space wavelength $\lambda$
with $a\sim \lambda/n_{\mathrm{med}}$.}
\label{fig:sketch}
\end{figure}

We use a surface modification scheme to alter the index of
refraction of the medium \cite{luopt07,huang:013824}.
An incident wave impinging on a smooth surface with incident angle
larger than the critical angle will be totally reflected.
However a proper surface grating will allow the wave to be transmitted.
This is equivalent to give the incident wave a transverse momentum kick.
In the case that the grating period is much smaller than the
incident wavelength, an effective refractive index $n_{\mathrm{eff}}$ can be
used to describe the refraction at the modified
surface. For a binary-staircase lens with a plano-concave shape
as shown in Fig. \ref{fig:sketch}, this effective index
is related to the bulk refractive index of the medium $n_{\mathrm{med}}$ by
\cite{luopt07}
\begin{equation}\label{eq:fresnel_zones}
n_{\mathrm{eff}} = n_{\mathrm{med}} -\dfrac{\lambda}{a}
\end{equation}
where `$a$' is a fixed step length along the optical axis (or surface periodicity) and $\lambda$ is the free space wavelength
(with $a<\lambda$). The number of steps $N_\mathrm{steps}$ or zones is then $R/a$, where $2R$ is the transverse size of the
binary-staircase lens. The focal length $f$ is calculated by using the formula $f=R/(1-n_{\mathrm{eff}})$. To obtain a good
focus, $a\sim\lambda/n_{\mathrm{med}}$ (Abbe's diffraction limit). In the present case, $\lambda=1550$ nm and $a$ was chosen as
450 nm with $n_{\mathrm{med}}$ $=$ 3.231 for the transverse electric (TE) modes and $n_{\mathrm{med}}$
 $=$ 3.216 for the transverse magnetic (TM) modes. $a$ has been chosen as an arbitrary value in
the vicinity of $\lambda/n_{\mathrm{med}}$.  $N_\mathrm{steps}$ $=$ 11, so that $2R$ reads 10 $\mu$m. Thus $n_\mathrm{eff}$ is
$-$0.2133 and $-$0.2889 for TE and TM modes, respectively.

\begin{figure}[htbp]
\centering{\includegraphics[width=7.5cm]{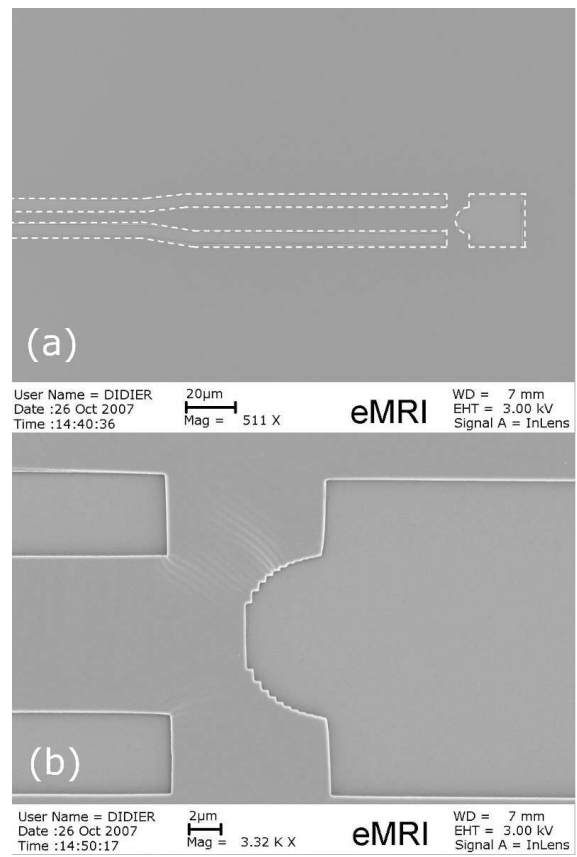}}
\caption{(a) Bird's eye view of the tapered waveguide and the binary-staircase lens,
(b) close-up view of the binary-staircase lens. }
\label{fig:overallview}
\end{figure}

The fabrication platform consisted of a 400 nm InGaAsP core layer on
an InP substrate with a 200 nm InP top cladding layer. The waves are
trapped and propagate within the core layer plane with an effective
permittivity of 3.231 (TE modes) and 3.216 (TM modes). The final
structure for optical measurements consisted of three sub units
(shown in Fig. \ref{fig:overallview}(a)): (i) a 0.5 mm long
waveguide, laterally tapered, having 5 $\mu$m wide trenches on
each side; The taper starts at a distance of 100 $\mu$m from the
edge of the waveguide, with a core width varying from 5 $\mu$m to
10 $\mu$m. (ii) Binary-staircase plano-concave lens with 10 zones
on the optical axis, having a step height of 450 nm and a transverse
size of 10 $\mu$m, located at a distance of 5 $\mu$m from the
tapered end of the waveguide (shown in Fig.
\ref{fig:overallview}(b)). (iii) And finally an open cavity
(semi-circle juxtaposed to a 20 $\mu$m $\times$ 20 $\mu$m
square) at the end of the binary-staircase lens.

An analogue structure, having the same geometrical
dimensions but bearing no steps (or zones), was fabricated. The
purpose of the analogous design was to prove that the periodicity of
the steps are decisive structure elements to realizing a
negative-index prototype. The structures were written using electron
beam lithography on polymethylmethacrylate (PMMA) resist. Pattern
transfers to a silicon nitride working mask and subsequently to the
InP/InGaAsP layers were achieved with a reactive ion etching (RIE)
method.

In the characterization experiment, a continuous wave tunable
semiconductor laser (1550 nm--1580 nm) was used as the input light
source. The laser light was coupled into the cleaved end of the
input waveguides using a monomode lensed fiber (working distance
$\approx$ 14$\mu$m and FWHM $\approx$ 2.5 $\mu$m in air) mounted
on a five-axis positioning stage. An infrared (IR) camera (Hamamatsu
Model C2741) connected to a microscope port aids the initial
alignment to optimize the IR light coupling from the optical fiber
to the waveguide. In the FDTD simulation, a 10 $\mu$m wide plane
parallel, Gaussian beam was chosen as incident field for the grating
lens. In the actual sample, the 5 $\mu$m wide input facet of the
waveguide was inversely tapered to 10 $\mu$m width (see Fig.
\ref{fig:overallview}(a)) so that the propagating Gaussian beam is
expanded sufficiently inside the guiding channel before reaching the
device end.  The planar wavefront after emerging from the
binary-staircase lens is expected to focus in the air
cavity.

\begin{figure}[htbp]
\centering{\includegraphics[width=8.3cm]{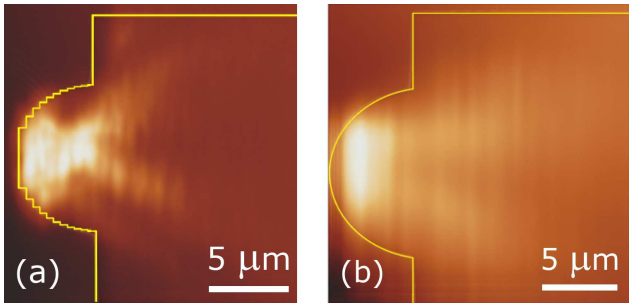}}
 \caption{(Color online) Optical images, from an optical scanning microscope,
obtained at $\lambda$ = 1550 nm around the focal point of (a) the binary-staircase lens,  (b)
   and the analogous structure with no zones/steps (semi-circle with smooth
   walls). Note that focusing is observed only with the binary-staircase lens.}
\label{fig:NSOM}
\end{figure}

A tapered fiber probe (250 nm aperture diameter) metallized with a thin chromium and gold layer was raster scanned just above the
sample surface. The output end of the fiber probe was connected to nitrogen cooled germanium detector (North Coast Scientific
Corp. Model $\#$ EO-817L). Additionally, a typical lock-in amplifier was utilized to optimize the detection scheme. Scanning the
fiber tip at a constant height about 500 nm above the sample surface allowed us to probe the optical intensity distribution over
a grid of $256\times256$ points spanning $15\times15$ $\mu$m$^2$ area. The reconstructed image is shown in Fig.
\ref{fig:NSOM}(a).

Intensity distribution near the cavity center clearly shows the
light focusing from the binary-staircase lens. Identical focusing
fingerprints were observed when the experiment was repeated over a
range of wavelengths varying from 1510 nm to 1580 nm. Another
controlled experiment was performed where the binary-staircase lens
was replaced by an analogous structure (having the same geometrical
features) with no steps. In the latter case, as shown in Fig.
\ref{fig:NSOM}(b), no beam focusing was observed. Nevertheless we
can distinguish a bright spot near the device's edge, which is
attributed to a sudden beam divergence as it propagates into open
space from initial confinement in the InGaAsP core waveguide layer
(diffraction).

\begin{figure}
\center{\includegraphics[width=8.7cm]{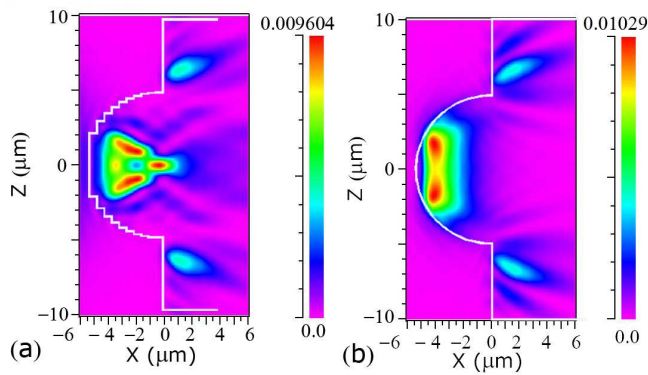}} \caption{(Color online) 3D FDTD simulations of (a) the plano-concave
binary-staircase lens, and (b) the lens having the same geometrical dimensions as the binary-staircase one, but bearing no steps
(or zones).} \label{fig:plc3d}
\end{figure}

The 3D FDTD simulations were performed using
perfectly matched layer boundary conditions that minimize
reflections at the edges. The chosen input field excitation for the
FDTD simulation was a TE polarized Gaussian beam which closely
resembles the beam shape of the fiber source in actual experiment.
The energy density of the propagating H-field was mapped at
different plane heights. Figures \ref{fig:plc3d}(a) and
\ref{fig:plc3d}(b) shows the simulated H-field density of the
binary-staircase lens and the analogue structure at about 800 nm
above the center of the core layer, respectively.

The traditional metamaterials structures are composed of arrays of split ring resonators and metal wires. This type of metallic
structures, which operates under resonance, becomes lossy at optical frequencies due to the inherent imaginary part of the
metal's permittivity. The purely dielectric system, such as the one mentioned in this letter, is free from these drawbacks and
thus has low intrinsic material loss, which is a clear-cut advantage for optical frequency operations. Extrinsic losses in the
binary-staircase dielectric structure itself arise solely from the imperfections in the fabrication (e.g. surface and sidewall
roughness).

We have experimentally designed a binary-staircase optical element
having an effective negative index of refraction, whereby the
surface periodicity of the structure acted as the tunable parameter
for controlling the sign change of the refractive index. The beam
propagation in the plano-concave lens was simulated using in-house
3D FDTD codes. Based on the design and
simulations, we have nano-engineered a prototype structure in an
InP/InGaAsP heterostructure tailored for the 1.55 $\mu$m
wavelength, where indium phosphide (InP) is a natural starting
fabrication platform for wholesale integration of passive and active
devices for a complete system-on-a-chip at this frequency.
Characterization of the prototype with a near-field scanning optical
microscope revealed that the plano-concave binary-staircase lens can
act as a convex lens and thereby focusing plane waves.  No focusing
is achieved if the zones are removed, reinforcing the fact the
steps are the decisive structure elements. A notable aspect of our
work is the extension of electromagnetic properties (that are
theoretically available) of optical elements for possible
integration in optoelectronic circuits.

This work was supported by the Air Force Research Laboratories,
Hanscom through grant no. FA8718-06-C-0045 and the National Science
Foundation through grant no. PHY-0457002. This work was performed in
part at the Kostas Center for High-Rate Nanomanufacturing at
Northeastern University, and Center for Nanoscale Systems, a member
of the National Nanotechnology Infrastructure Network, under NSF
award no. ECS-0335765.



\end{document}